\documentclass[nohyper,12pt,letterpaper]{JHEP3}
\usepackage{epsfig}
\usepackage[latin1]{inputenc}
\usepackage{bbm,amsfonts}
\usepackage{graphicx}
\usepackage{amssymb,amsmath}
\usepackage{comment}

\author{
  Marco S. Bianchi$^\ast$,
  Matias Leoni$^{\dag}$
  and Silvia Penati$^{\ddag}$ \\\\
  $^\ast$Departamento de F\'isica, Universidad de Oviedo, 33007, Oviedo, Spain
  \\\\
  $^{\dag}$Physics Department, FCEyN-UBA \& IFIBA-CONICET\
  
  Ciudad Universitaria, Pabell\'on I, 1428, Buenos Aires, Argentina
  \\\\
  $^{\ddag}$Dipartimento di Fisica, Universit\`a di Milano--Bicocca and
  INFN, Sezione di Milano--Bicocca, Piazza della Scienza 3, I-20126 Milano, Italy
  \qquad\\\\
  E-mail: \email{bianchimarco@uniovi.es, leoni@df.uba.ar,
    silvia.penati@mib.infn.it
  }
}

\abstract{In planar ${\cal N} = 4$ SYM we study a particular class of helicity preserving amplitudes. These are scalar amplitudes whose flavor configuration is chosen in such a way that only a limited number of diagrams is allowed, which exhibit an iterative structure. For such amplitudes we evaluate the tree level and one--loop contributions, providing a general formula valid for any number of particles. The ratio between the one--loop and tree level results is a simple combination of dual conformally invariant box functions with at most two massive legs. Along with the MHV and NMHV series, this constitutes the third known infinite sequence of one--loop amplitudes in ${\cal N}=4$ SYM.}

\preprint{ August 2012 \\ FPAUO-12/09}

\title{Minimally helicity violating, maximally simple scalar amplitudes in $\mathcal{N}=4$ SYM}


\csname @addtoreset\endcsname{equation}{section}

\def\bseq{\begin{subequation}}  
\def\eseq{\end{subequation}}
\def\bsea{\begin{subeqnarray}}  
\def\esea{\end{subeqnarray}}

\newcommand{\beq}{\begin{equation}}
\newcommand{\bea}{\begin{eqnarray}}
\newcommand{\eea}{\end{eqnarray}}
\newcommand{\eeq}{\end{equation}}

\newcommand {\non}{\nonumber}

\renewcommand{\a}{\alpha}
\newcommand{\da}{\dot \alpha}
\renewcommand{\b}{\beta}
\newcommand{\db}{\dot \beta}

\newcommand{\dg}{\dot \gamma}
\renewcommand{\d}{\delta}

\newcommand{\pa}{\partial}
\newcommand{\g}{\gamma}
\newcommand{\G}{\Gamma}

\newcommand{\e}{\epsilon}

\renewcommand{\l}{\lambda}

\newcommand{\F}{\Phi}

\newcommand{\p}{\pi}

\newcommand{\Db}{\overline{D}}

\newcommand{\Fb}{\overline{\F}}

\newcommand{\adot}{\dot{\alpha}}

\newcommand{\thb}{\overline{\theta}}

\renewcommand{\thb}{\overline{\theta}}

\def\Mb{\kern 2pt\mathchoice
        {
         \vbox{\hrule width10pt height 0.4pt depth 0pt
         \kern 1.2pt\hbox{\kern -2pt$\displaystyle M$}}}
        {
         \vbox{\hrule width10pt height 0.4pt depth 0pt
         \kern 1.2pt\hbox{\kern -2pt$\textstyle M$}}}
        {
\vbox{\hrule width6pt height 0.4pt depth 0pt
         \kern 1.0pt\hbox{\kern -2pt$\scriptstyle M$}}}
        {
         \vbox{\hrule width5pt height 0.4pt depth 0pt
         \kern 0.8pt\hbox{\kern -2pt$\scriptscriptstyle M$}}}}

\def\Sb{\kern 2pt\mathchoice
        {
         \vbox{\hrule width6pt height 0.4pt depth 0pt
         \kern 1.2pt\hbox{\kern -2pt$\displaystyle S$}}}
        {
         \vbox{\hrule width6pt height 0.4pt depth 0pt
         \kern 1.2pt\hbox{\kern -2pt$\textstyle S$}}}
        {
         \vbox{\hrule width3.5pt height 0.4pt depth 0pt
         \kern 1.0pt\hbox{\kern -2pt$\scriptstyle S$}}}
        {
         \vbox{\hrule width3pt height 0.4pt depth 0pt
         \kern 0.8pt\hbox{\kern -2pt$\scriptscriptstyle S$}}}}

\def\Rb{\kern 2pt\mathchoice
        {
         \vbox{\hrule width5.5pt height 0.4pt depth 0pt
         \kern 1.2pt\hbox{\kern -2.5pt$\displaystyle R$}}}
        {
         \vbox{\hrule width5.5pt height 0.4pt depth 0pt
         \kern 1.2pt\hbox{\kern -2.5pt$\textstyle R$}}}
        {
         \vbox{\hrule width3.5pt height 0.4pt depth 0pt
         \kern 1.0pt\hbox{\kern -2.2pt$\scriptstyle R$}}}
        {
         \vbox{\hrule width3pt height 0.4pt depth 0pt
         \kern 0.8pt\hbox{\kern -2.2pt$\scriptscriptstyle R$}}}}

  \def\pp{{\mathchoice
      {
          \kern 1pt%
          \raise 1pt
          \vbox{\hrule width5pt height0.4pt depth0pt
            \kern -2pt
            \hbox{\kern 2.3pt
              \vrule width0.4pt height6pt depth0pt
              }
            \kern -2pt
            \hrule width5pt height0.4pt depth0pt}%
            \kern 1pt
       }
        {
          \kern 1pt%
          \raise 1pt
          \vbox{\hrule width4.3pt height0.4pt depth0pt
            \kern -1.8pt
            \hbox{\kern 1.95pt
              \vrule width0.4pt height5.4pt depth0pt
              }
            \kern -1.8pt
            \hrule width4.3pt height0.4pt depth0pt}%
            \kern 1pt
        }
        {
          \kern 0.5pt%
          \raise 1pt
          \vbox{\hrule width4.0pt height0.3pt depth0pt
            \kern -1.9pt  
            \hbox{\kern 1.85pt
              \vrule width0.3pt height5.7pt depth0pt
              }
            \kern -1.9pt
            \hrule width4.0pt height0.3pt depth0pt}%
            \kern 0.5pt
        }
        {
          \kern 0.5pt%
          \raise 1pt
          \vbox{\hrule width3.6pt height0.3pt depth0pt
            \kern -1.5pt
            \hbox{\kern 1.65pt
              \vrule width0.3pt height4.5pt depth0pt
              }
            \kern -1.5pt
            \hrule width3.6pt height0.3pt depth0pt}%
            \kern 0.5pt
        }
    }}

  \def\mm{{\mathchoice
               {
                 \kern 1pt
           \raise 1pt    \vbox{\hrule width5pt height0.4pt depth0pt
                  \kern 2pt
                  \hrule width5pt height0.4pt depth0pt}
                 \kern 1pt}
               {
                \kern 1pt
           \raise 1pt \vbox{\hrule width4.3pt height0.4pt depth0pt
                  \kern 1.8pt
                  \hrule width4.3pt height0.4pt depth0pt}
                 \kern 1pt}
               {
                \kern 0.5pt
           \raise 1pt
                \vbox{\hrule width4.0pt height0.3pt depth0pt
                  \kern 1.9pt
                  \hrule width4.0pt height0.3pt depth0pt}
                \kern 1pt}
               {
               \kern 0.5pt
         \raise 1pt  \vbox{\hrule width3.6pt height0.3pt depth0pt
                  \kern 1.5pt
                  \hrule width3.6pt height0.3pt depth0pt}
               \kern 0.5pt}
               }}

\def\pd{{\kern0.5pt
           + \kern-5.05pt \raise5.8pt\hbox{$\textstyle.$}\kern
0.5pt}}

\def\pmd{{\kern0.5pt
          \pm \kern-5.05pt
\raise6.3pt\hbox{$\textstyle.$}\kern1.5pt}}

\def\md{{\mathchoice
   {
      {{\kern 1pt - \kern-6.2pt \raise5pt\hbox{$\textstyle.$}\kern
1pt}}}
    {
      {{\kern 1pt - \kern-6.2pt \raise5pt\hbox{$\textstyle.$}\kern
1pt}}}
    {
      {\kern0.5pt - \kern-5.05pt
\raise3.4pt\hbox{$\textstyle.$}\kern0.5pt}}
    {
      {\kern0.5pt - \kern-5.05pt
\raise3.4pt\hbox{$\textstyle.$}\kern0.5pt}}}}

\def\beq{\begin{equation}}
\def\eeq{\end{equation}}
\def\bea{\begin{eqnarray}}
\def\eea{\end{eqnarray}}
\def\Tr{\text{Tr}}
\def\tr{\text{tr}}
\def\a{\alpha}
\def\b{\beta}
\def\g{\gamma}

\def\d{\delta}
\def\e{\epsilon}

\def\th{\theta}
\def\l{\lambda}
\def\G{\Gamma}

\def\F{\Phi}

\begin{document}

\section{Introduction}

The evaluation of scattering amplitudes in interacting gauge theories is one of the main tasks in high--energy particle physics. So far, the most far--reaching results have been obtained for maximally supersymmetric 
Yang--Mills theory in four dimensions (for a recent review and a quite exhaustive list of references, see for instance \cite{Dixon:2011xs}). The development of new efficient computational techniques, the use of general recursion relations and the exploitation of the high degree of symmetry of the planar sector of the theory have allowed the evaluation of scattering amplitudes well beyond the lowest perturbative order for processes involving a small number of particles and the determination of compact formulae for complete series of amplitudes at tree and one loop level.

Up to one loop, a general expression for MHV $n$--point gluon amplitudes has been given in \cite{Bern:1994zx, Bern:1994cg}, whereas a complete result for NMHV $n$--point gluon amplitudes can be found in \cite{Georgiou:2004by, Bern:2004bt}. General results for NMHV amplitudes involving gluinos and scalars have been derived in \cite{Bidder:2005in, Risager:2005ke}. For split--helicity gluon amplitudes a general expression valid for any number of particles has been determined at tree level \cite{Britto:2005dg}. 

${\cal N}=4$ supersymmetry can be exploited for constructing a superamplitude \cite{Nair:1988bq, Witten:2003nn, Drummond:2008vq}, which in its expansion in the superspace Grassmannian variables contains all the component amplitudes corresponding to all possible configurations of external particles (gluons, fermions and scalars) and all possible configurations of helicities.
So far, the complete $n$--point superamplitude has been constructed at tree level \cite{Drummond:2008cr} and results at one--loop exist for MHV \cite{Witten:2003nn, Brandhuber:2004yw} and NMHV \cite{Drummond:2008vq, Drummond:2008bq} $n$--points superamplitudes. Beyond one loop, integrands for MHV and NMHV $n$--points superamplitudes have been given in \cite{ArkaniHamed:2010kv,ArkaniHamed:2010gh}.

In this paper we determine a new general iterating formula for one loop corrections to a particular class of color ordered amplitudes. These are purely scalar amplitudes where the external chiralities and flavors are chosen in such a way that at tree level only one diagram contributes, so making its direct computation very easy (this was first noticed in \cite{AHPS}).  Using ${\cal N}=1$ superspace formalism, this simplicity carries over at loop level where the few diagrams contributing can be obtained from the tree level one by repeated insertion of simple building blocks. This allows for a direct determination of the general one--loop correction for any number of external scalar particles. The general result is given in eq. (\ref{final}) in terms of one--mass, two--mass easy and two--mass hard scalar box functions. Along with the MHV and NMHV series, this constitutes the third known infinite series of one--loop amplitudes in ${\cal N}=4$ SYM.

The plan of the paper is as follows: in Section \ref{sec:2} we introduce the class of simple scalar amplitudes, reinterpreting them in a manifestly $\mathcal{N}=1$ supersymmetric setup. In Section \ref{sec:3} we rederive the tree level results of \cite{AHPS} in our formalism. In Section \ref{sec:4} we present the new computation at one loop. We first compute explicitly the simplest four and six point amplitudes, showing how the well known expression for the corresponding $\mathcal{N}=4$ superamplitudes may be rediscovered from this different perspective. Then we generalize the result to an arbitrary number of external scalars and perform a non--trivial check of its correctness by comparing its IR behavior with the expected universal structure.

\section{The simplest scalar amplitudes}\label{sec:2}

We consider the scattering of $n$ scalar particles in ${\cal N}=4$ SYM theory, in the large $N$ limit. In the following we always understand color ordering and deal with partial amplitudes.

The theory contains six real scalar fields $\phi_i$ transforming in the vector representation of the $SO(6)_R$ R--symmetry group and in the adjoint representation of the gauge group $SU(N)$. They interact with the gluons of the gauge sector, with the four fermions through Yukawa interactions, and among themselves by a quartic scalar potential. Hence scattering processes involve in principle many contributions from all these interactions, making their traditional perturbative evaluation quite unfeasible.

However, as noticed in \cite{AHPS}, there exist a particular class of scalar amplitudes that at tree level receive contributions from a single Feynman diagram.
These are amplitudes where the order of the external scalar particles has been chosen in such a way that no two adjacent ones share the same $SO(6)_R$ index. The only planar tree level diagram contributing to this color ordered amplitude is just a chain of scalar quartic vertices, as shown in Figure \ref{fig:treecomponents}, and then it is very easy to compute.
\FIGURE{
  \centering
\includegraphics[width = 1. \textwidth]{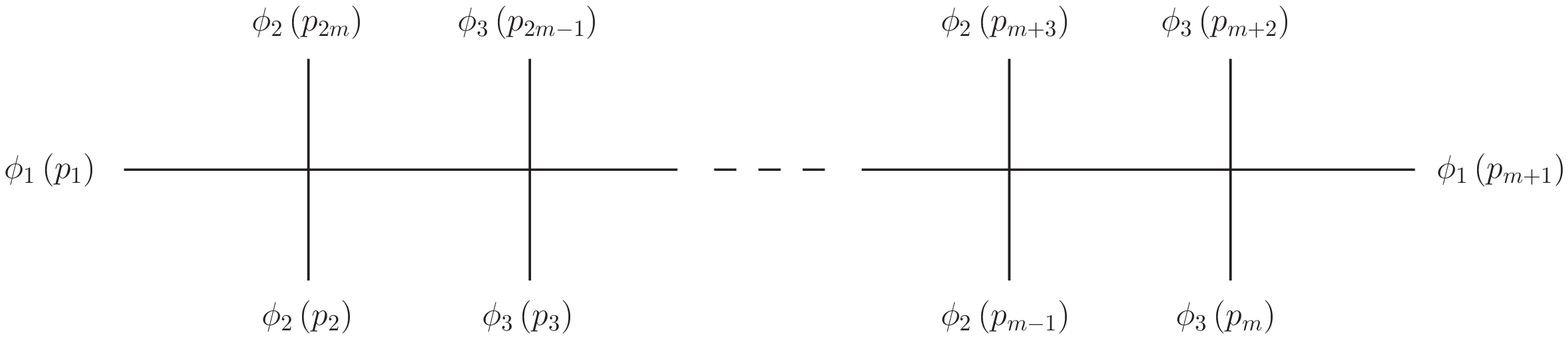}
\caption{Example of scalar amplitude at tree level in component formalism.}
\label{fig:treecomponents}
}

In a manifestly $\mathcal{N}=4$ superspace formalism, on--shell states can be organized in a single chiral superfield, according to the following expansion in powers of the four Grassmannian coordinates $\eta^A$
\begin{eqnarray} \label{fitinto}
  \Phi(p,\eta) &=& G^{+}(p) + \eta^A\, \Gamma_A(p) + \frac{1}{2}\,\eta^A\, \eta^B\, S_{AB}(p)
  + \frac{1}{3!}\,\eta^A\,\eta^B\,\eta^C\, \e_{ABCD}\, \bar\Gamma^{D}(p) \non \\
  &&\  + \frac{1}{4!}\,\eta^A\,\eta^B\,\eta^C\, \eta^D\, \e_{ABCD}\, G^{-}(p)
\end{eqnarray}
Here, $G^\pm(p)$ are the two helicity states for gluons and $\G_A, \bar{\G}^A$ the two fermionic states. Introducing $\tilde{S}^{AB} \equiv \frac12\, \e^{ABCD} S_{CD}$ the scalars satisfy the reality condition 
$\overline{ (S _{AB})} = \tilde{S}^{AB}$, 
and the components of the $S_{AB}, \tilde{S}^{AB}$ antisymmetric tensors are identified with the scalars carrying $SO(6)$ index, according to
\bea
 && S_{1,i+1} = \phi_i - i \, \phi_{i+3} \non \\
 && \tilde{S}^{1,i+1} = \phi_i + i \, \phi_{i+3} \qquad , \quad i =1,2,3
\eea
In this language, an $n$--point superamplitude can be constructed from the vacuum expectation value of a string of $n$  $\Phi(p, \eta)$ superfields. On--shell conservation laws, ordinary superconformal invariance and dual superconformal covariance constrain it to be of the form \cite{Drummond:2008vq}
\begin{equation}\label{expnonmhv2}
    \mathcal{A}(\Phi_1,\ldots,\Phi_n) = \mathcal{A}_n^{\rm MHV} \mathcal{P}_n
\end{equation}
where ${\cal P}_n$ is a polynomial in the $SU(4)_R$ singlet $\eta^4 \equiv \frac{1}{4!} \e_{ABCD} \eta^A \eta^B \eta^C \eta^D$, which is dual superconformal invariant. We write
\begin{equation}
\label{expansion}
\mathcal{P}_n = 1 + \mathcal{P}_n^{\rm NMHV} + \mathcal{P}_n^{\rm N^2MHV} + \,\, \ldots \,\,
\mathcal{P}_n^{\rm N^kMHV} + \,\, \ldots \,\, + \mathcal{P}_n^{\overline{\rm MHV}}
\end{equation}
where each term is homogeneous in $\eta^4$ with  ${\rm deg}(\mathcal{P}_n^{\rm N^kMHV}) = 4{\rm k}$ and ${\rm k}=0, \cdots, n-4$.  According to the particular value of k we obtain a different kind of ratio function
$\mathcal{A}_n^{\rm N^kMHV} /\mathcal{A}_n^{\rm MHV}$, ranging from the MHV to the MHV-conjugate cases.

We remind that at tree level 
\beq
\label{tree}
\mathcal{A}_{n, tree}^{\rm MHV} = 
\frac{\delta^4(p) \delta^8(q)}{\langle12\rangle \ldots \langle n1\rangle } 
\eeq

From this general construction it is easy to realize that purely scalar amplitudes only occur for an even number of external particles and they always correspond to helicity--preserving (or "minimally violating") amplitudes. In fact, since each scalar component  $S_{AB}$ is associated to a pair of Grassmannian variables, in order to ensure $SU(4)$ invariance, it has to enter the amplitude together with its conjugate $\tilde{S}^{AB}$. This gives rise to amplitudes with the same number of $S$ and $\tilde{S}$ fields.

As a consequence, the spectrum of purely scalar amplitudes does not fill the whole expansion (\ref{expansion}). For fixed $n\equiv 2m$, a purely scalar amplitude will appear only in $\mathcal{P}_{2m}^{{\rm N}^{(m-2)} {\rm MHV}}$.  This corresponds to having a MHV amplitude at four points, a NMHV amplitude at six points,
a ${\rm N^2MHV}$ at eight points and so on and so forth, according to a pattern that strictly resembles the one of three--dimensional ABJM--type theories \cite{BLM}.

\vskip 20pt

In order to evaluate perturbatively scattering amplitudes, it is worth using $\mathcal{N}=1$ superspace
formalism.

We embed the scalar fields into 3 complex chiral superfields $X$, $Y$ and $Z$, whose dynamics together with the one for the vector superfield, is described by the action (\ref{eq:action}).
In terms of the scalars carrying a $SO(6)$ index and the ones in the antisymmetric representation of $SU(4)$, we make the specific identification
\bea
\label{fields}
X \equiv \phi_1 + i \phi_4 = \tilde{S}^{12}     \quad &,& \quad \bar{X} \equiv \phi_1 - i \phi_4 = S_{12}
\non \\
Y \equiv \phi_2 + i \phi_5 = \tilde{S}^{13}     \quad &,& \quad \bar{Y} \equiv \phi_2 - i \phi_5 = S_{13}
\non \\
Z \equiv \phi_3 + i \phi_6 = \tilde{S}^{14}     \quad &,& \quad \bar{Z} \equiv \phi_3 - i \phi_6 = S_{14}
\eea
${\cal N}=1$ superamplitudes will be extracted from contributions to the effective action corresponding to strings of chiral and antichiral superfields.  

We choose to concentrate on the particular set of superamplitudes
\beq\label{eq:ampsuper}
{\cal A}_{2m}(Y_1 X_2 \cdots X_{m}\, \bar Y_{m+1} \bar X_{m+2} \cdots \bar X_{2m})
\eeq
which at tree level receive a single planar contribution corresponding to a string of cubic superpotential vertices (see Figure \ref{fig:tree}). The absence of other typologies of planar diagrams is due to the fact that no adjacent superfields appear with the same flavor and opposite chirality. This is the way the condition for having a single diagram at tree level works in ${\cal N}=1$ language. 
Other alternative choices of flavors are related to this one by $SU(3)_R$ R-symmetry transformations, so we can focus on these particular superamplitudes without loosing generality.

\FIGURE{
  \centering
\includegraphics[width = 1. \textwidth]{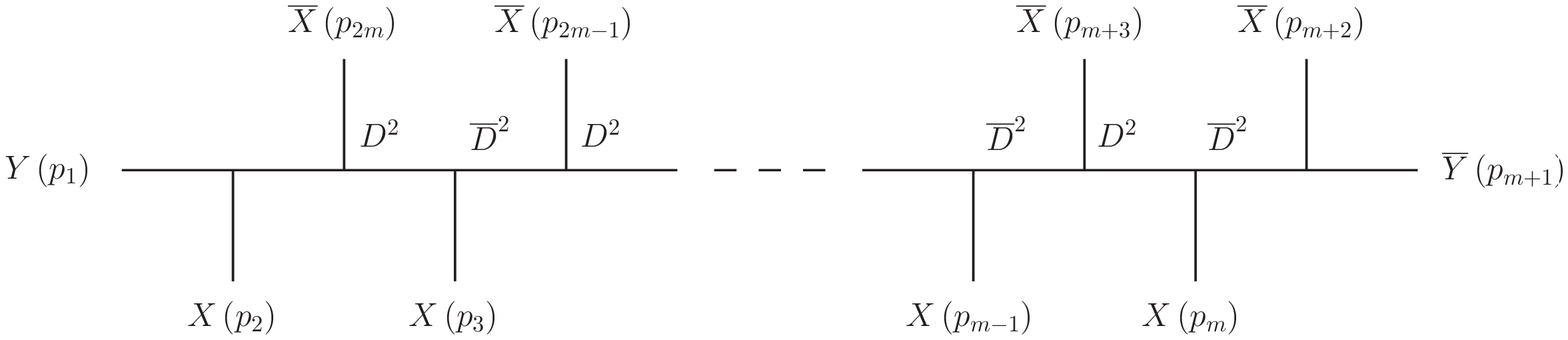}
\caption{The single diagram contributing to the tree-level amplitude in ${\cal N}=1$ formalism.}
\label{fig:tree}
}

Superamplitudes (\ref{eq:ampsuper}) belong to the class of the so-called split--helicity amplitudes.
They are invariant under reflection
\begin{equation}
{\cal A}_{2m}(Y_1 \cdots X_{m}\, \bar Y_{m+1} \cdots \bar X_{2m})  = {\cal A}_{2m}(\bar X_{2m} \cdots \bar Y_{m+1}\, X_{m} \cdots Y_1)
\end{equation}
and under parity transformations, which exchange chiral with antichiral superfields
\begin{equation}
{\cal A}_{2m}(Y_1 \cdots X_{m}\, \bar Y_{m+1} \cdots \bar X_{2m})  = {\cal A}_{2m}(\bar Y_1 \cdots \bar X_{m}\, Y_{m+1} \cdots X_{2m})
\end{equation}
By direct inspection of the symmetries of the corresponding Feynman diagrams they turn out to be invariant also under the following $\mathbb{Z}_2$ symmetries
\bea
&& {\cal A}_{2m}(Y_1\, X_2 \cdots X_{m}\, \bar Y_{m+1}\, \bar X_{m+2} \cdots \bar X_{2m})  = {\cal A}_{2m}(Y_1\, X_{2m} \cdots X_{m+2}\, \bar Y_{m+1}\, \bar X_{m} \cdots \bar X_{2})
\non \\
\non \\
&& {\cal A}_{2m}(Y_1\, X_2 \cdots X_{m}\, \bar Y_{m+1}\, \bar X_{m+2} \cdots \bar X_{2m})  = {\cal A}_{2m}(Y_{m+1}\, X_{m} \cdots X_{2}\, \bar Y_{1}\, \bar X_{2m} \cdots \bar X_{m+2}) \non \\
\eea

In the next Sections we perform the one--loop evaluation of these amplitudes for an arbitrary number of external particles. Using ${\cal N}=1$ superspace formalism has the advantage to involve a smaller number of diagrams compared to the calculation in components. Moreover, from the final expression of (\ref{eq:ampsuper}) we could extract not only the purely scalar amplitude we are interested in, but other sets of amplitudes involving matter fermions.

The same strategy of computing amplitudes that at tree level involve only superpotential interactions was successfully  used in \cite{BLMPS1,BLMPS2}  for determining diagrammatically the two--loop correction to the four point superamplitude in ABJ(M) and in \cite{BLMPS3}  for the six--point superamplitude at one loop.

\section{Tree level}\label{sec:3}

Given the particular configuration of the $n=2m$ external fields (\ref{eq:ampsuper}), at tree level and in the planar limit these ${\cal N}=1$ superamplitudes receive one single contribution corresponding to the diagram depicted in Figure \ref{fig:tree}.

The result is easily worked out by performing the D--algebra on the supergraph. This amounts to integrating by parts the spinorial derivatives on the external fields in order to obtain an expression local in the spinorial variables. This results in several terms which can be collected in the following compact expression
\beq
\G^{(0)}_{2m} =
\frac{(ig)^{2m-2}}{P}\,\!\! \int\!\! d^4\, \th\,  Y (p_1)\, X (p_2)  \left( \prod_{i=1}^{m-2}
\bar X(p_{2m-i+1})\, D^2\, X(p_{i+2})\, \Db^2 \right) \bar Y(p_{m+1})\, \bar X(p_{m+2})
\eeq
where $P$ comes from the product of the propagators (see Appendix B for notations on momentum invariants)
\beq
P = p^2_{m+1,m+2}\, \prod_{i=1}^{m-2}\, p^2_{2m-i+2;2i}\, p^2_{2m-i+1;1+2i}
\eeq
Integrating on the $\th$-variable, distributing the spinorial derivatives in all possible ways on the superfields and  introducing polarization spinors for fermions we may obtain all the component amplitudes.

The purely scalar component is extracted by applying the spinorial derivatives in such a way that they always appear in even number on a given superfield. Using the on--shell conditions
\beq
D^2 X= \bar{D}^2 \bar{X} = 0  \qquad , \qquad \bar{D}_{\adot} D_\a X(p) = p_{\a\adot}\, X(p)  \qquad , \qquad D_\a \bar{D}_{\adot} \bar{X}(p) =
p_{\a\adot}\, \bar{X}(p)
\eeq
and similarly for $Y$,  all spinorial derivatives get converted to external momenta and give rise to a numerator expressed in terms of momentum invariants
\beq
\prod_{i=1}^{m-1}\, \left(-p^2_{2-i;2i}\right)
\eeq
Labeling the internal momenta in Figure \ref{fig:tree} from the leftmost chiral vertex up to its analogue at the opposite end, it is easy to see that  this product cancels all internal propagators at odd sites, leaving the final expression for the tree level scalar amplitude
\beq\label{eq:2mtree}
{\cal A}^{(0)}_{2m} = g^{2m-2}\, \prod_{i=1}^{m-2}\, \frac{1}{p^2_{2m-i+1;\, 2i+1}}
\eeq
that agrees with the result found in \cite{AHPS}.

The simplest case corresponds to $m=2$, which gives a constant for the tree--level scattering of four scalar particles. Comparing with the Parke--Taylor gluon amplitude, the MHV denominator $\prod_{i=1}^4\, \langle i,i+1 \rangle$  is not present, as can be ascertained solving super Ward identities.
Alternatively, this can be understood by extracting the particular scalar amplitude we are considering from the general expression of the 4pt ${\cal N}=4$ superamplitude (\ref{tree}) where ${\cal P}_4=1$. According to the field identification (\ref{fields}), our 4pt scalar amplitude corresponds to the $\eta_1^4 \eta_1^2 \eta_2^3 \eta_2^4 \eta_3^1 \eta_3^3 \eta_4^1 \eta_4^2$ component. Extracting this term from the fermionic delta function
$\d^{(8)}(q)= \frac{1}{2^4}\, \prod_{A=1}^{4}\, \sum_{i,j} \, \langle i,j \rangle\, \eta^A_i\, \eta_{j\,A}$ it is immediate to recognize the emergence of a factor $- \langle 12 \rangle \langle 23 \rangle \langle 34 \rangle \langle 41 \rangle$, which cancels  the MHV denominator in the superamplitude, so leading to a constant.

For $m=3$ we obtain the amplitude for six scalars
\beq
\label{6pt}
{\cal A}^{(0)}_{6} = \frac{g^4}{p_{6;3}^2}
\eeq
Again, we checked that this coincides with the $\eta_1^4 \eta_1^2 \eta_2^3 \eta_2^4 \eta_3^3 \eta_3^4 \eta_4^1 \eta_4^3 \eta_5^1 \eta_5^2\eta_6^1 \eta_6^2$ component of the ${\cal N}=4$ superamplitude (\ref{tree}) with ${\cal P}_6 = {\cal P}_{6,tree}^{\rm NMHV}$ \cite{Drummond:2008vq}. In the notations of \cite{Drummond:2008vq} this component receives contributions only from the $R_{1;36}$ R--invariant (see eq. (6.17) in that paper).

\section{One--loop}\label{sec:4}

In this section we compute the one--loop correction to the process (\ref{eq:ampsuper}). In the planar limit all one--loop diagrams are order $\l \equiv g^2N$ compared to the tree level counterpart. We evaluate the ratio ${\cal M}^{(1)}_{2m}$ defined as
\beq
{\cal A}_{2m} = {\cal A}_{2m}^{(0)} \left( 1 + \l \, c_{\G}\, {\cal M}^{(1)}_{2m} + {\cal O}\left( \l^2 \right) \right)
\eeq
where $c_{\G}$ is the customary factor as defined in Appendix B (see eq. \ref{rG}).

In the large $N$ limit, and taking into account that one--loop corrections to chiral propagators vanish, the relevant supergraphs are obtained from the tree level one by adding one vector propagator joining two chiral lines in all possible planar ways. The corrections can be schematically drawn as in Figure \ref{fig:blocs1L}.
\FIGURE{
  \centering
\includegraphics[scale = 0.6]{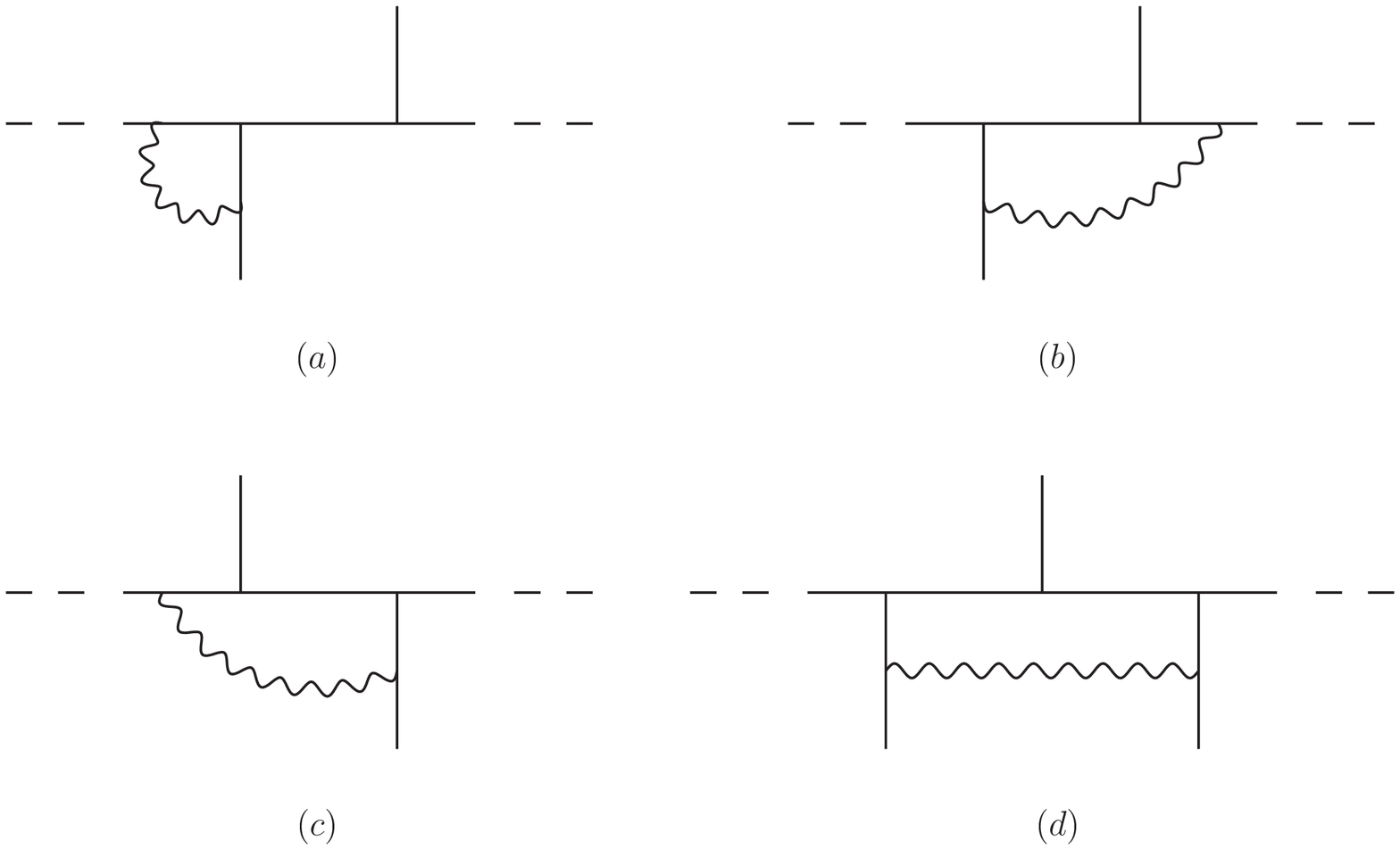}
\caption{Building--block diagrams contributing at one--loop.}
\label{fig:blocs1L}
}
We sketch the calculation for the 4pt and 6pt amplitudes before concentrating on the general $n$pt case.

\subsection{Four--points and six--points}

At one loop, the scattering of four chiral superfields
\beq
{\cal A}_{4}(Y(1) X(2)  \, \bar Y(3) \bar X(4))
\eeq
involves only the two supergraphs shown in Figure \ref{fig:4pt1L}.

\FIGURE{
  \centering
\includegraphics[scale = 0.6]{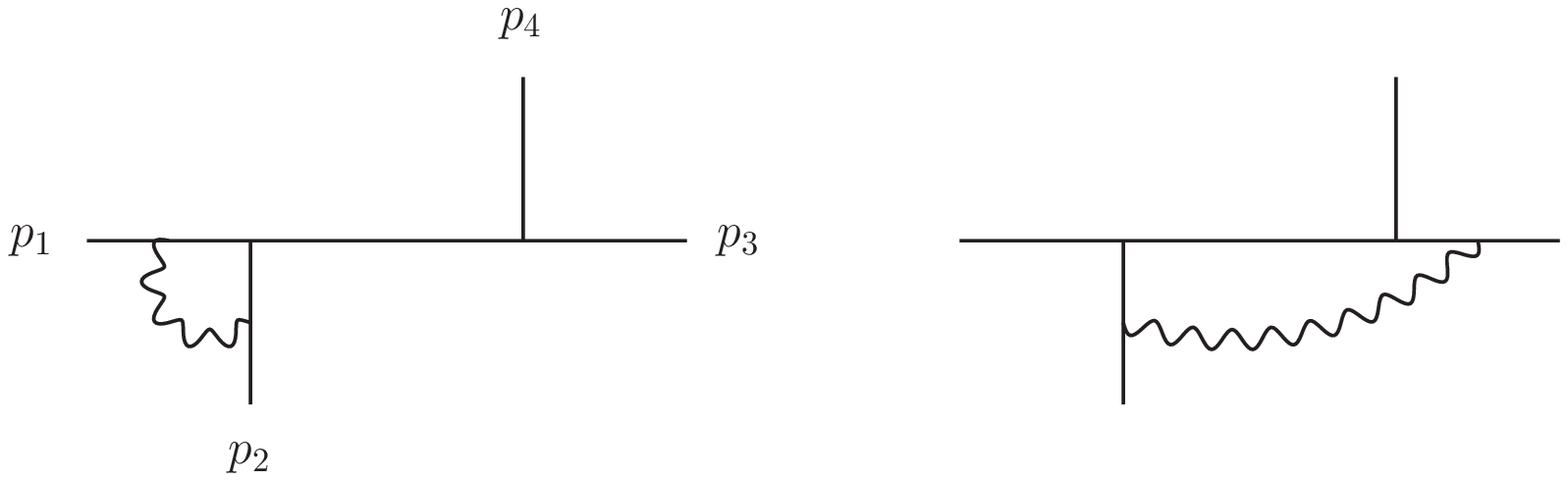}
\caption{The two types of diagram contributing to the one--loop four--point amplitude. Two more are obtained by acting with parity transformation on these ones.}
\label{fig:4pt1L}
}

Performing the D--algebra they give rise to ordinary momentum integrals corresponding to triangle and box diagrams. We regularize IR divergences by dimensional regularization.  The notations and the results for the integrals are listed in the Appendix \ref{app:integrals}, and follow \cite{Bern:1994cg}, up to a different normalization factor.

Focusing on the purely scalar component and taking into account permutations and overall factors (we omit only an overall $1/(4\pi)^{2-\e}$ which we restore in the final result), the contributions from the two diagrams read
\bea
(a) &=& - 2\, s\, I^{1\mathrm{m}}_{3;1} \times {\cal A}_4^{(0)} \non\\
(b) &=& \left[ s\, I^{1\mathrm{m}}_{3;1} + p_4^{\a\da}\, p_{1\,\b\da}\, p_2^{\b\db}\,
\int \frac{d^{4-2\e} k}{(2\p)^{4-2\e}}\, \frac{k_{\a\db}}{k^2(k+p_4)^2(k+p_{41})^2(k+p_{4;3})^2}
~ + \right.\non\\&& ~
+ \left( p_1 \,\leftrightarrow\, p_3\,,\, p_2 \,\leftrightarrow\, p_4  \right) \bigg]
\,\times\, {\cal A}_4^{(0)}
\eea
Using trace formulae (\ref{app:traces}), they sum up to
\beq
{\cal M}_4^{(1)} = - \frac{s\,t}{(4\pi)^{2-\e}}\, I_{4}^{0\mathrm{m}}
\eeq
As expected, the contributions corresponding to triangle integrals cancel against opposite contributions from the box--like ones, leaving a massless box, according to the well--known result for the 4pt gluon amplitude \cite{Ellis:1985er}--\cite{Bern:1991aq}.

\vskip 20pt

At six points we want to compute the amplitude
\beq
{\cal A}_{6}(Y(1) X(2) X(3) \, \bar Y(4) \bar X(5) \bar X(6))
\eeq
The actual contribution to the six--point ${\cal N}=1$ superamplitude is given by diagrams in Fig. \ref{fig:6pt1L} plus their parity duals, obtained by exchanging $p_i \leftrightarrow p_{3+i}$. The diagrams have been grouped according to  the number of propagators flowing inside the loop. Compared to the four--point case, a new kind of diagram appears, which formally leads to a pentagon integral (see Fig. \ref{fig:6pt1L}$(g)$).

\FIGURE{
  \centering
\includegraphics[scale = 0.58]{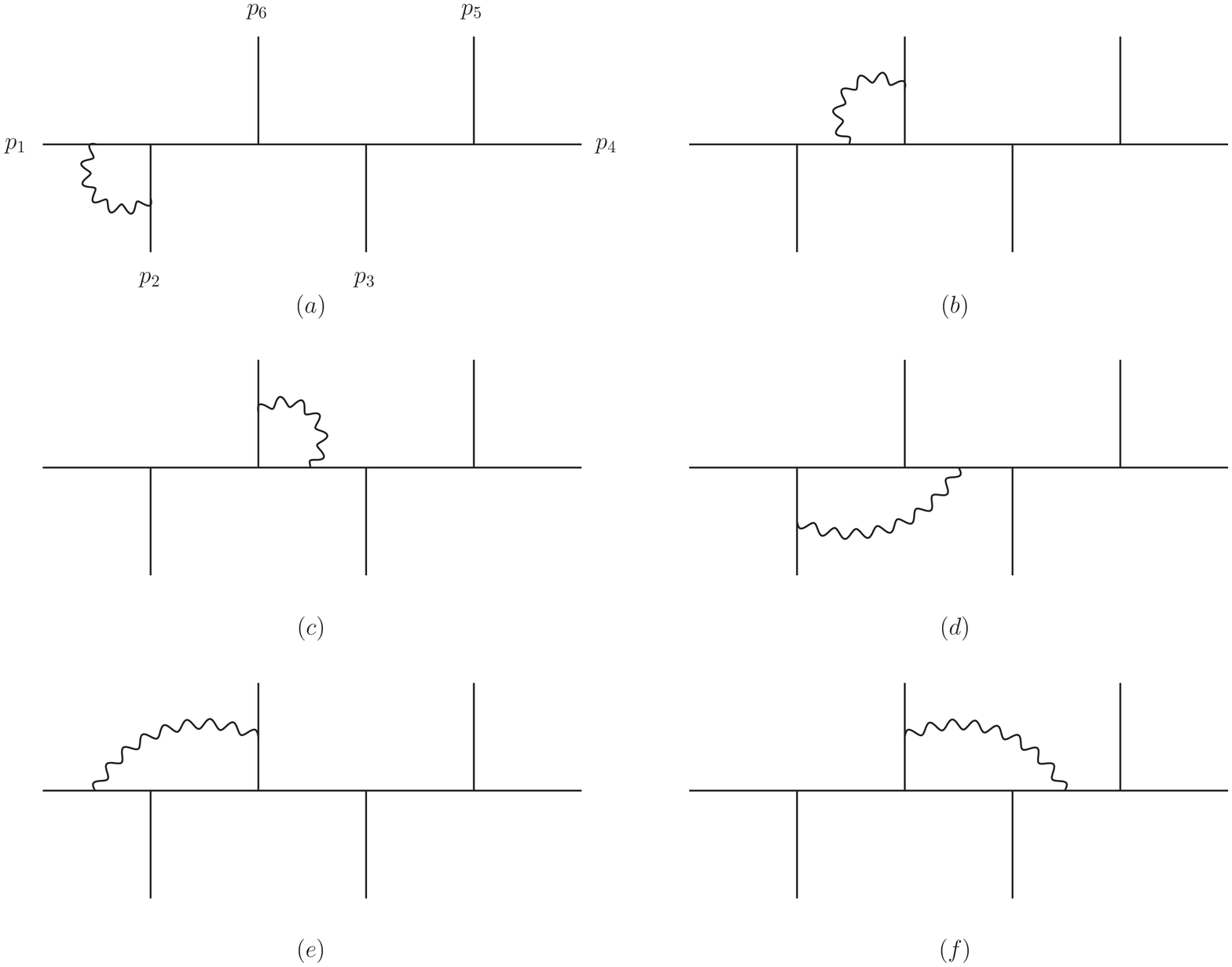}
\includegraphics[scale = 0.58]{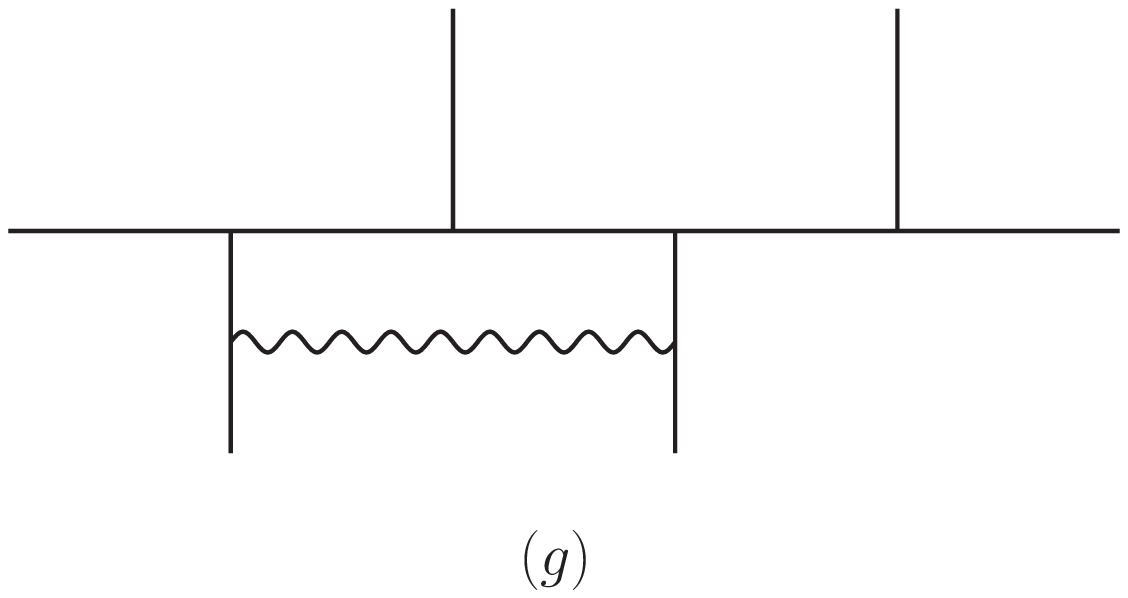}
\caption{Types of diagrams contributing to the six--point one--loop amplitude. Seven more diagrams are obtained by acting with parity transformation on these ones.}
\label{fig:6pt1L}
}

After performing D--algebra, graphs of the type \ref{fig:6pt1L}$(a)$,$(b)$,$(c)$ give rise to triangle--like integrals.  The first one and its dual correspond to one--mass triangle integrals, whereas the others give rise to two--mass triangle integrals. Precisely,
\beq
(a) =  -\frac{p_{12}^2}{p_{6;3}^2}\, I_{3:\, 1}^{1\mathrm{m}} \quad,\quad
(b) = - I_{3:\,2;1}^{2\mathrm{m}} \quad,\quad
(c) = -\frac{p_{12}^2}{p_{6;3}^2}\, I_{3:\,2;1}^{2\mathrm{m}}
\eeq
and similarly for the dual ones.

Concerning supergraphs with four loop propagators, the D--algebra works differently for diagrams \ref{fig:6pt1L}$(d)$,$(e)$ compared to diagram \ref{fig:6pt1L}$(f)$. In all cases we obtain a triangle and a vector box--like integral, but given the different configuration of spinorial derivatives that survive on the external fields, in the former two diagrams both integrals contribute to the scalar part of the amplitude, whereas in the latter one only the term proportional to the triangle integral survives. Precisely, these diagrams give
\bea
(d) & = & \frac{p_{12}^2}{p_{6;3}^2}\, I_{3:\, 1}^{1\mathrm{m}} + \frac{1}{p_{6;3}^2}\, \int \frac{d^{4-2\e} k}{(2\p)^{4-2\e}}\, \frac{\Tr \left( p_6\, p_1\, p_2\, k \right)}{k^2 (k+p_{6})^2 (k+p_{16})^2 (k+p_{6;3})^2} \non\\
(e) & = & \frac{p_{12}^2}{p_{6;3}^2}\, I_{3:\,2;1}^{2\mathrm{m}} - \frac{1}{p_{6;3}^2}\, \int \frac{d^{4-2\e} k}{(2\p)^{4-2\e}}\, \frac{\Tr \left( p_6\, p_1\, p_2\, k \right)}{k^2 (k-p_1)^2 (k-p_{61})^2 (k+p_2)^2}
 \non\\
(f) & = & I_{3:\,2;1}^{2\mathrm{m}}
\eea
A similar result holds for their parity duals.

If we now sum diagrams \ref{fig:6pt1L}$(a)$--\ref{fig:6pt1L}$(f)$ it is easy to realize that the triangle integrals cancel pairwise, while for the two vector box--like integrals a suitable change of integration variables allows them to be paired. Working out the traces at numerator we eventually obtain a scalar box integral with one massive leg
\beq
\label{a-f}
(a)+(b)+(c)+(d)+(e)+(f)=
- \frac{p_{12}^2\, p_{16}^2}{p_{6;3}^2}\, I_{4:\,3}^{1\mathrm{m}}
\eeq
Finally, there are the new diagrams with five loop propagators as in Figure \ref{fig:6pt1L}$(g)$, which, nevertheless, can be easily solved. The nice outcome of the D--algebra decomposition is that only two--mass hard box integrals contribute to the scalar component of the amplitude, e.g.
\beq
\label{g}
(g) = - p_{23}^2\, I_{4:\,2;4}^{2\mathrm{m}\, h}
\eeq
Combining (\ref{a-f}), (\ref{g}) and their parity duals, inserting the appropriate color and combinatorial factors and employing the dual conformally invariant \cite{Drummond:2006rz} scalar box functions $F$ \cite{Bern:1994cg} defined in (\ref{F}), the one--loop correction at six points divided by the tree level amplitude (\ref{6pt}) reads
\bea\label{eq:sixpoint}
{\cal M}_6^{(1)} &=&
2\,
F_{6:\,3}^{1\mathrm{m}}
+2\,
F_{6:\,6}^{1\mathrm{m}}
+2\,
F_{6:\,2;4}^{2\mathrm{m}\, h}
+2\,
F_{6:\,2;1}^{2\mathrm{m}\, h}
\eea
The involved integrals have been computed long ago and an explicit evaluation of (\ref{eq:sixpoint}) can be found for instance in \cite{Drummond:2008vq}.

The expression (\ref{eq:sixpoint}) is perfectly consistent with the result for the NMHV six--point gluon amplitude found in \cite{Bern:1994cg}, where using unitarity cuts it was proved that the coefficient of the two--mass easy box integral, which might potentially be present, is actually zero. Given the constraints imposed on the superamplitude by dual superconformal invariance, our procedure provides a simple diagrammatic explanation of that finding, being it an immediate consequence of the structure of the supergraphs and the way D--algebra works on them. In fact, the only diagrams in Figure \ref{fig:6pt1L} which might give rise to a two--mass easy box integral are diagrams $(f)$ and $(g)$, but the corresponding superfield configurations arising from D--algebra never produce purely scalar terms.

At six points, the one--loop ${\cal P}_6^{\rm NMHV}$ ratio in (\ref{expansion})  can be expressed as a sum over R--invariants, dressed by a function $V^{}$ of the three conformal cross--ratios \cite{Drummond:2008vq} (see also \cite{Dixon:2011nj}).
Our result (\ref{eq:sixpoint}) divided by ${\cal P}_6^{\rm MHV}$ coincides with the function $V_6^{(3)}$ correcting the $R_{1;36}$ invariant at one loop (see eq. (5.26) of Ref. \cite{Drummond:2008vq}).  

For the two simple cases of  MHV four--point and NMHV six--point amplitudes, one could have determined the exact expression of the ratios $\mathcal{P}_4^{\rm MHV}$ and $\mathcal{P}_6^{\rm NMHV}$ at one loop simply computing the corresponding purely scalar amplitudes and combining the result with the dual superconformally invariant ansatz for the superamplitude.
Unfortunately, this is not possible any longer starting from the eight--point ${\rm N^2MHV}$ amplitudes, as we will highlight in the next Section.

\subsection{$n$--points} 

The experience gained in the evaluation of the previous simple cases, can be used for generalizing the one--loop calculation to $n=2m$ external scalar particles.
This can be easily accomplished, as no new typologies of diagrams emerge compared to the six--point case.

The most efficient way to perform the calculation is to evaluate building blocks in Figure \ref{fig:blocs1L} and sum over all possible block insertions inside a $2m$--leg diagram. 

\FIGURE{
  \centering
\includegraphics[width = \textwidth]{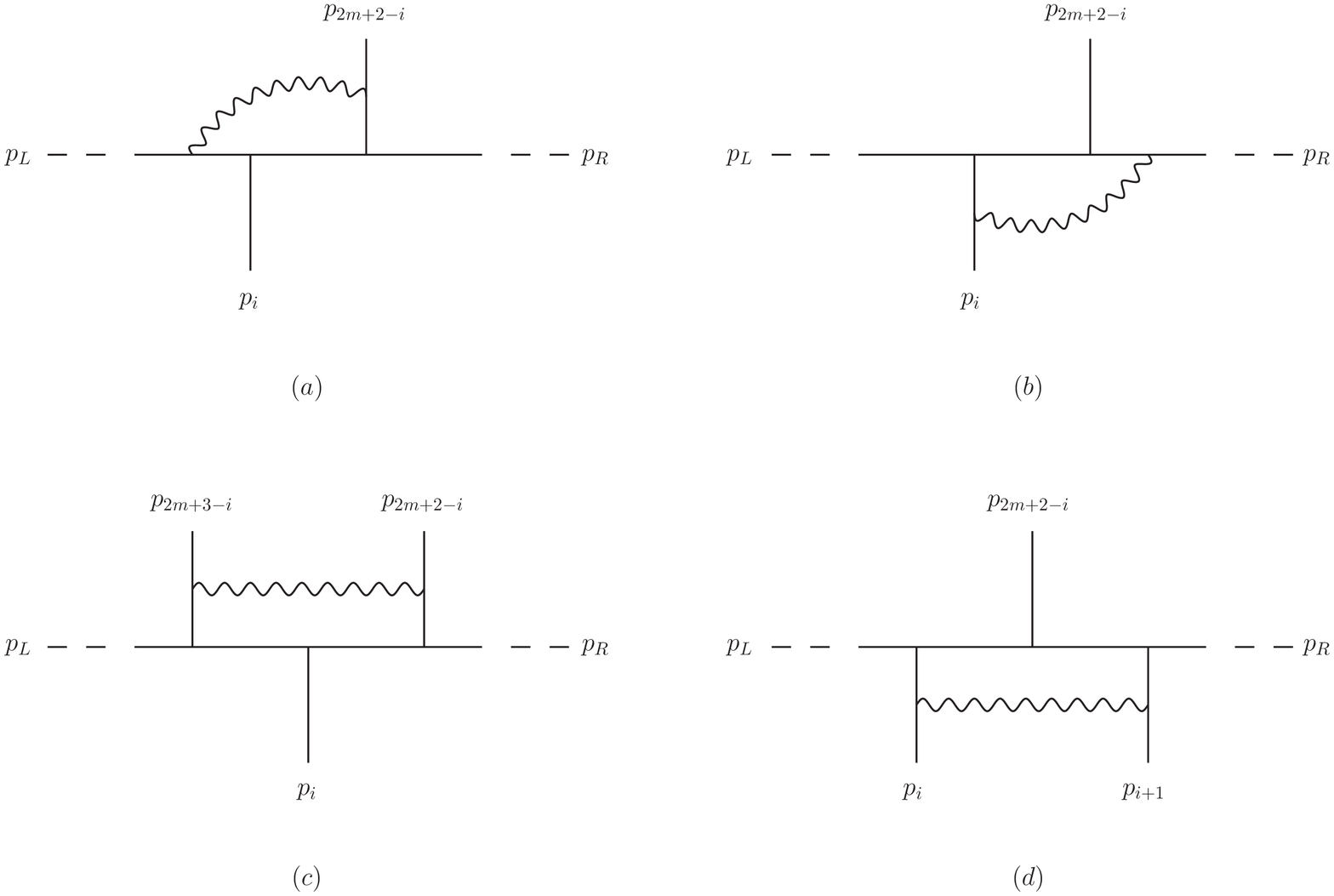}
\caption{Relevant diagrams contributing to the $2m$--point one--loop scalar amplitude.}
\label{fig:2mpoints}
}
 
The expected cancellation of triangle integrals suggests the way to conveniently group the diagrams in order to have triangles disappearing already at an intermediate stage. 
From a case by case analysis it is immediate to realize that triangle integrals from triangle--like diagrams \ref{fig:blocs1L}$(a)$ and those coming from the D--algebra reduction of box--like ones \ref{fig:blocs1L}$(b)$ cancel pairwise, leaving boxes only. As for six--point case,  building block \ref{fig:2mpoints}$(c)$ turns out to contribute to the purely scalar component of the amplitude only with a triangle--like integral, which cancels a contribution from a genuine triangle diagram. In conclusion, only building block diagrams in Figure \ref{fig:2mpoints} truly contribute with box--like integrals. 

From the insertion of blocks \ref{fig:blocs1L}$(a)$ and \ref{fig:blocs1L}$(b)$ inside the tree level diagram, vector two--mass easy box integrals are obtained
\bea
(a) &=& + {\cal A}_{2m}^{(0)}\times \int \frac{d^{4-2\e} k}{(2\p)^{4-2\e}}\, \frac{\Tr \left( p_{2m+2-i}\, p_{2m+3-i;2i-3}\, p_i\, k \right)}{k^2 (k+p_{2m+2-i})^2 (k+p_{2m+2-i;2i-2})^2 (k+p_{2m+2-i;2i-1})^2}
\non\\
(b) &=& - {\cal A}_{2m}^{(0)}\times \int \frac{d^{4-2\e} k}{(2\p)^{4-2\e}}\, \frac{\Tr \left( p_{2m+2-i}\, p_{2m+3-i;2i-3}\, p_i\, k \right)}{k^2 (k-p_{2m+3-i;2i-3})^2 (k-p_{2m+2-i;2i-2})^2 (k+p_i)^2}
\eea
For any value $i=3, \cdots , m-1$, for $m\geq 4$, summing these contributions the vector part cancels and we are left with a scalar two--mass easy box 
\beq
(a)+(b) = - \left( p_{2m+2-i;2i-2}^2\, p_{2m+3-i;2i-2}^2 - p_{2m+3-i;2i-3}^2\, p_{i+1;2m-2i+1}^2 \right)\, {\cal A}_{2m}^{(0)}\, I_{4:\,2m-2i+1;i+1}^{2\mathrm{m}e}
\eeq
When the insertion of blocks $(a)$ and $(b)$ happens at the edges of the tree level graph, that is $i=2$ and $i= m$ in Figure \ref{fig:2mpoints}, one more external momentum becomes massless and we obtain one--mass  integrals. For instance, for $i=2$ we have
\bea
(a)\big|_{i=2} &=& + {\cal A}_{2m}^{(0)}\times \int \frac{d^{4-2\e} k}{(2\p)^{4-2\e}}\, \frac{\Tr \left( p_{2m}\, p_1\, p_2\, k \right)}{k^2 (k+p_{2m})^2 (k+p_{2m,1})^2 (k+p_{2m;3})^2}
\non\\
(b)\big|_{i=2} &=&  -{\cal A}_{2m}^{(0)}\times \int \frac{d^{4-2\e} k}{(2\p)^{4-2\e}}\, \frac{\Tr \left( p_{2m}\, p_1\, p_2\, k \right)}{k^2 (k-p_1)^2 (k-p_{2m,1})^2 (k+p_2)^2}
\eea
As above, these combine pairwise leaving the one--mass scalar box contribution
\beq
(a)+(b) = - p_{12}^2\, p_{2m,1}^2\, {\cal A}_{2m}^{(0)}\, I_{4:\,3}^{1\mathrm{m}}
\eeq
The two analogous diagrams at the opposite corner are worked out in the same fashion.

In addition, there are other two--mass hard scalar box integrals coming from blocks \ref{fig:2mpoints}$(c)$ and \ref{fig:2mpoints}$(d)$, which have to be summed over the insertion leg $i$.
Explicitly, after a straightforward generalization of how to work out D--algebra and extract the scalar part (which is spelled out in Appendix \ref{app:pentagon}), the corresponding contributions to the purely scalar component read
\bea\label{eq:pentagons}
(c) &=& - p_{i,i+1}^2\,\, p_{i+1;2m-2i+1}^2\,\, {\cal A}_{2m}^{(0)}\times I_{4:\,2m-2i;i+2}^{2\mathrm{m}h}
\non\\
(d) &=& - p_{2m-i+2,2m-i+3}^2\,\, p_{2m-i+3;2i-3}^2\,\, {\cal A}_{2m}^{(0)}\times I_{4:\,2i-4;2m-i+4}^{2\mathrm{m}h}
\eea

Summing up all contributions with the relative factors, dividing by the tree level amplitude (\ref{eq:2mtree}) and expressing the integrals in terms of the $F$ scalar box functions, the final answer reads
\bea\label{eq:result}
{\cal M}_{2m}^{(1)}\, &=& \,
2\,
F_{2m:\,3}^{1\mathrm{m}} 
\,\,+\,\, 2\,
F_{2m:\,m+3}^{1\mathrm{m}} 
\,\,+\,\,2\, \sum_{i=3}^{m-1}\, F_{2m:\,2m-2i+1\,;\,i+1}^{2\mathrm{m}e} \,\, 
\non\\ &&
+ \,\,2\, \sum_{i=3}^{m}\, 
F_{2m:\,2i-4\,;\,2m-i+4}^{2\mathrm{m}h} 
\,\,+\,\,2\, \sum_{i=2}^{m-1}\, 
F_{2m:\,2m-2i\,;\,i+2}^{2\mathrm{m}h}
\eea
Including the one--mass box functions into the sum of the two--mass easy, and relabeling indices in the second sum, we can rewrite the ratio ${\cal M}_{2m}^{(1)}$ as
\beq
\label{final}
\boxed {
{\cal M}_{2m}^{(1)}\, = \,
\,2\, \sum_{i=2}^{m}\, F_{2m:\,2m-2i+1\,;\,i+1}^{2\mathrm{m}e} \,\, +
\,\,2\, \sum_{i=2}^{m-1}\, \left( \,
F_{2m:\,2m-2i\,;\,i+2+m}^{2\mathrm{m}h}
\,\,+\,\,
F_{2m:\,2m-2i\,;\,i+2}^{2\mathrm{m}h} \, \right)
}
\eeq
This is the main result of the paper. As explained above, the sums are performed over the insertion position of the blocks in Figure \ref{fig:2mpoints}, translating into different momenta entering the legs of the scalar box functions.
Again these box diagrams are only one--mass and two--mass easy and hard, and the summation over the massless external legs is a straightforward generalization of the six--point result. Dual conformal invariance, as well as parity invariance, are manifest in (\ref{final}).

The occurrence of box integrals with at most two massive legs makes it evident that this set of scalar amplitudes is not sufficient for determining the whole ${\rm N}^{m-2}{\rm MHV}$ superamplitude. In fact, already at eight points, ${\rm N^2MHV}$ gluonic amplitudes contain four--mass box integrals \cite{Britto:2004nc}, which are not present in the scalar sector.  This is consistent with the obvious expectation that for $m$ increasing more and more components will be needed to fix $2m$--point superamplitudes.  
Our one--loop findings may offer one constraint for determining the complete ${\rm N}^{m-2}{\rm MHV}$ one--loop correction for $m>3$.

\subsection{Infrared behavior}

As a check of the correctness of result (\ref{eq:result}) we can test whether it reproduces the expected structure of IR divergences, which in euclidean signature reads \cite{Giele:1991vf}--\cite{Catani:1999ss}
\beq\label{eq:ir}
{\cal M}^{(1)}_{2m}\big|_{IR} = - \frac{1}{\e^2}\, \sum_{i=1}^{2m} \left(\frac{\mu^2}{p^2_{i,i+1}}\right)^{\e}  
\eeq
By using the known results  for box integrals (a list may be found in \cite{Bern:1994zx}) we extract the following divergent terms
\bea
F_{2m:\,3}^{1\mathrm{m}}\big|_{IR} &=& - \frac{1}{\e^2} \left[
\left( p^2_{2m;2} \right)^{-\e} + \left( p^2_{1;2} \right)^{-\e} - \left( p^2_{2m;3} \right)^{-\e}
\right]\non\\
F_{2m:\,m+3}^{1\mathrm{m}} \big|_{IR} &=& - \frac{1}{\e^2} \left[
\left( p^2_{m;2} \right)^{-\e} + \left( p^2_{m+1;2} \right)^{-\e} - \left( p^2_{m;3} \right)^{-\e}
\right]\non\\
F_{2m:\,2m-2i+1;i+1}^{2\mathrm{m}e}\big|_{IR} &=& - \frac{1}{\e^2} \left[
\left( p^2_{i;2m-2i+2} \right)^{-\e} + \left( p^2_{i+1;2m-2i+2} \right)^{-\e} \right.\non\\&& \left. ~~~~ - \left( p^2_{i+1;2m-2i+1} \right)^{-\e} - \left( p^2_{2m-i+3;2i-3} \right)^{-\e}
\right]
\non\\
F_{2m:\,2i-4;2m-i+4}^{2\mathrm{m}h}\big|_{IR}  &=& - \frac{1}{\e^2} \left[
\frac12\, \left( p^2_{2m-i+2;2} \right)^{-\e} + \left( p^2_{2m-i+3;2i-3} \right)^{-\e} \right.\non\\&& \left. ~~~~ -\frac12\, \left( p^2_{2m-i+4;2i-4} \right)^{-\e} -\frac12\, \left( p^2_{i;2m-2i+2} \right)^{-\e}
\right]
\non\\
F_{2m:\,2m-2i;i+2}^{2\mathrm{m}h}\big|_{IR} &=& - \frac{1}{\e^2} \left[
\frac12\, \left( p^2_{i;2} \right)^{-\e} + \left( p^2_{i+1;2m-2i+1} \right)^{-\e} \right.\non\\&& \left. ~~~~ -\frac12\, \left( p^2_{i+2;2m-2i} \right)^{-\e} -\frac12\, \left( p^2_{2m-i+2;2i-2} \right)^{-\e}
\right]
\eea
Summing these terms according to the prescription (\ref{eq:result}) we can straightforwardly ascertain that (\ref{eq:ir}) is recovered. This provides a strong consistency check.

Turning the logic around, the result (\ref{final}) could have been derived from the knowledge of the universal IR behavior of the amplitudes and some intuition from the diagrammatic expansion of the purely scalar one loop amplitudes.  
In fact, looking at the topology of graphs which contribute at one loop it is immediate to realize that box--like  diagrams with three and four massive legs never arise.  One--mass integrals emerge from vector corrections \ref{fig:2mpoints}$(a)$  and \ref{fig:2mpoints}$(b)$ at the two extrema of the chain; two--mass easy integrals  
arise when these corrections are internal; two--mass hard boxes come from vector insertions of the form \ref{fig:2mpoints}$(c)$  and \ref{fig:2mpoints}$(d)$ which leave two $X$ fields or two $\bar X$ fields  as massless legs, respectively. Given the particular structure of the diagrams, which are obtained by iterative insertion of building blocks, it is reasonable to expect that all contributions associated to the insertion of a particular block will have the same coefficient. Moreover, parity invariance forces the coefficients of the two one--mass integrals to be the same, as well as the ones corresponding to blocks \ref{fig:2mpoints}$(c)$ and \ref{fig:2mpoints}$(d)$. We are then left with three unknown coefficients which can be determined from the request to have the correct $1/\e^2$ pole (this fixes the relative coefficient between the one--mass and the two--mass hard integrals) and the correct $1/\e$ pole (this fixes the relative coefficient of the two--mass easy integrals).

\vskip 25pt
\section*{Acknowledgements}
\noindent
M. Leoni thanks G. Giribet and J. Maldacena for an interesting discussion. This work has been supported in part by INFN and MIUR, by the research grants MICINN-09-FPA2009- 07122 and MEC-DGI-CSD2007-00042 and by the research project CONICET PIP0396.

\vfill
\newpage

\appendix

\section{Notations and conventions}\label{app:notations}

We work in four dimensional euclidean ${\cal N}=1$ superspace described by coordinates $(x^\mu, \th^\a,\, \thb^{\da})$, $\a, \da =1,2$. We follow conventions of \cite{superspace}. 

Superspace covariant derivatives are defined as
\begin{equation}
  D_\a = \pa_\a + \frac{i}{2}\,  \thb^{\da}\,  \pa_{\a\da}
  \qquad , \qquad \Db_{\da} = \bar\pa_{\da}
  + \frac{i}{2}\,  \th^\a\,  \pa_{\a\da}
\end{equation}
and satisfy the anticommutator
\begin{equation}
\label{algebra}
  \{D_\a ,\, \Db_{\da}\} = i\,  \pa_{\a\da}
\end{equation}

Given the algebra of Dirac $\left(\g^{\mu}\right)^\a{}_{\da}$ matrices  
\begin{eqnarray}
  (\gamma^\mu)^{\a\da}\, (\gamma^\nu)_{\a{}\da}
  =2\, g^{\mu\nu}
\end{eqnarray}
trace identities needed for loop calculations can be easily obtained  
\begin{eqnarray}\label{app:traces}
  &\tr(\gamma^\mu\, \gamma^\nu)
  &= -(\gamma^\mu)^{\a\da}\, (\gamma^\nu)_{\a\da}
  = -2\,  g^{\mu\nu} \\
  &\tr(\gamma^\mu\, \gamma^\nu\, \gamma^\rho\, \gamma^\sigma) &= (\gamma^\mu)^{\a\da}\, (\gamma^\nu)_{\b\da}\,
  (\gamma^\rho)^{\b\dg}\, (\gamma^\sigma)_{\a\dg} = \non\\
  &&= 2\, (g^{\mu\nu}\, g^{\rho\sigma}-g^{\mu\rho}\, g^{\nu\sigma}+g^{\mu\sigma}\, g^{\nu\rho}) - 2\, \epsilon^{\mu\nu\rho\sigma}
\end{eqnarray}
 
The $SU(N)$ generators $T^A$ ($A=1,\ldots, N^2-1$) are a set of
$N\times N$ hermitian matrices satisfying 
\beq
\left[T^A,\,T^B\right] = i\, f^{ABC}\, T^C
\eeq
They are normalized as $\Tr( T^A T^B )= \delta^{AB}$.

The ${\cal N}=1$ superspace action of ${\cal N}=4$ SYM reads
\bea\label{eq:action}
S &=& \int d^4x\, \left(\frac{1}{g^2}\, \int d\th^2\, \Tr \, W^2 +  \int d\th^4\, \Tr  (e^{-gV}\, \Fb_i\, e^{gV}\, \F^i)  \right.
\non\\&& \left.
+ \, i\, g\, \int d\th^2\, \Tr (\left[ X,\, Y \right]\, Z )+ i\, g\, \int d\thb^2\, \Tr (\left[ \bar X,\, \bar Y \right]\, \bar Z) \right)
\eea
where $\Phi_i=\{X,Y,Z\}$ and $W_\a = i\, \Db^2\, \left( e^{-gV}\, D_{\a}\, e^{gV} \right)$.

The propagators of the gauge and chiral superfields are:
\bea
\langle V(\th_1)V(\th_2) \rangle = -\frac{1}{p^2}\, \d^4(\th_1-\th_2)\quad,\quad
\langle \F(\th_1)\F(\th_2) \rangle = \frac{1}{p^2}\, \d^4(\th_1-\th_2)
\eea
and the interaction vertices relevant for the one--loop computation read
\bea
& - g\, f^{ABC}\, X^A\, Y^B\, Z^C \quad,\quad
- g\, f^{ABC}\, \bar X^A\, \bar Y^B\, \bar Z^C&\non\\&
i\, g\, f^{ABC}\, \F^{i\,A} \Fb_i^B\, V^C&
\eea

\section{Integrals}\label{app:integrals}

We define sums of adjacent momenta by
\beq
p_{i,j} = p_i + p_j
\eeq
and
\beq
p_{i;r} = \sum_{k=i}^{i+r-1}\, p_k
\eeq
Momentum integrals are regularized by dimensional regularization $D=4-2\e$.
The notation for relevant triangle and box integrals is as in \cite{Bern:1994cg}, up to a different normalization, due to euclidean signature.
\begin{itemize}
\item
Triangle integrals
\begin{itemize}
\item one--mass
\beq\label{app:tri1}
I^{1\mathrm{m}}_{3:i} = (4\pi)^{2-\e}\,
\int \frac{d^D k}{(2\p)^D}\, \frac{1}{k^2 (k-p_i)^2 (k+p_{i+1})^2}
\eeq
\item two--mass
\beq\label{app:tri2}
I^{2\mathrm{m}}_{3:r;i} = (4\pi)^{2-\e}\,
\int \frac{d^D k}{(2\p)^D}\, \frac{1}{k^2 (k-p_{i-1})^2 (k+p_{i;r})^2}
\eeq
\end{itemize}
\item
Box integrals
\begin{itemize}
\item one--mass
\beq\label{app:box1m}
I^{1\mathrm{m}}_{4:i}
= (4\pi)^{2-\e}\,
\int \frac{d^D k}{(2\p)^D}\, \frac{1}{k^2 (k+p_{i-3})^2 (k+p_{i-3;2})^2 (k+p_{i-3;3})^2}
\eeq
\item two--mass easy
\beq\label{app:box2me}
I^{2\mathrm{m}e}_{4:r;i}
= (4\pi)^{2-\e}\,
\int \frac{d^D k}{(2\p)^D}\, \frac{1}{k^2 (k+p_{i-1})^2 (k+p_{i-1;r+1})^2 (k+p_{i-1;r+2})^2}
\eeq
\item two--mass hard
\beq\label{app:box2mh}
I^{2\mathrm{m}h}_{4:r;i}
= (4\pi)^{2-\e}\,
\int \frac{d^D k}{(2\p)^D}\, \frac{1}{k^2 (k+p_{i-2})^2 (k+p_{i-2;2})^2 (k+p_{i-2;2+r})^2}
\eeq
\end{itemize}
\end{itemize}
For dual conformally invariant box integrals the box functions $F$ of \cite{Bern:1994cg} are employed, again with a different normalization
\begin{align}
\label{F}
F^{1\mathrm{m}}_{2m:i} & = - \frac{1}{2\, r_{\G}}
p_{i-2;2}^2\, p_{i-3;2}^2\, I^{1\mathrm{m}}_{4:i} \non\\
F^{2\mathrm{m}e}_{2m:r;i} & = - \frac{1}{2\, r_{\G}}
\left( p_{i-1;r+1}^2\, p_{i;r+1}^2 - p_{i;r}^2\, p_{i+r+1;n-r-2}^2 \right)\, I^{2\mathrm{m}e}_{4:r;i} \non\\
F^{2\mathrm{m}h}_{2m:r;i} & = - \frac{1}{2\, r_{\G}}
p_{i-2;2}^2\, p_{i-1;r+1}^2\, I^{2\mathrm{m}h}_{4:r;i}
\end{align}
Following the literature, $r_{\G}$ is defined as
\beq
\label{rG}
r_{\G} = \frac{\G(1+\e)\, \G(1-\e)^2}{\G(1-2\e)}
\eeq
and the one--loop amplitude is rescaled by an overall $c_{\G} = \frac{r_{\G}}{(4\pi)^{2-\e}}$ factor.

\section{Pentagon diagram D--algebra}\label{app:pentagon}

As a non--trivial example of D--algebra reduction, in this Appendix we consider the generic pentagon diagram of Figure \ref{fig:blocs1L}$(c)$ and spell out its D--algebra, leading to the result (\ref{eq:pentagons}).
The computation may be undergone diagrammatically as shown in Figure \ref{fig:pentagon_D-al}.
\FIGURE{
  \centering
\includegraphics[width = 1. \textwidth]{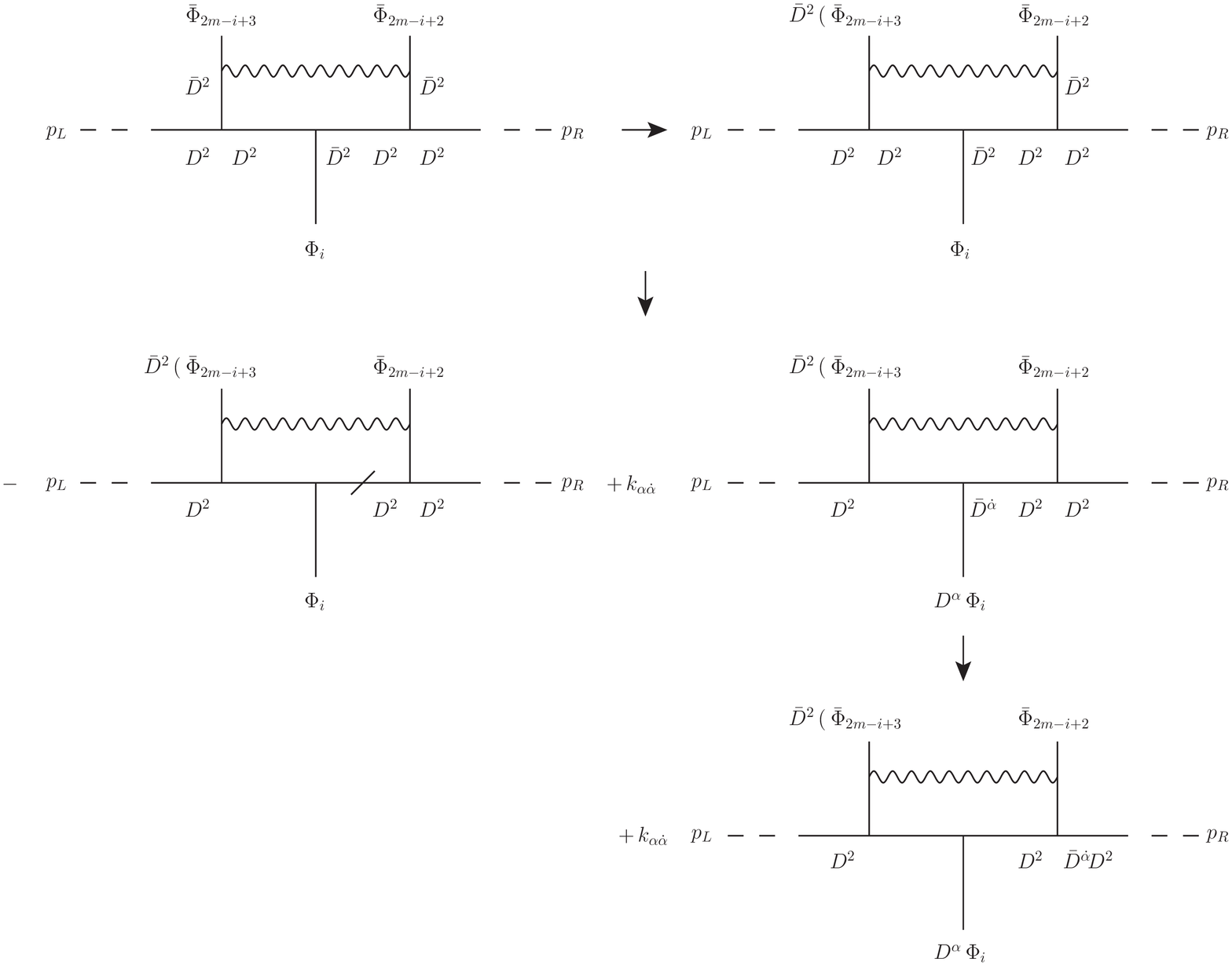}
\caption{D--algebra of the pentagon diagrams.}
\label{fig:pentagon_D-al}
}
The first term corresponds to a scalar box integral.
The second term does not contribute to the completely scalar component of the superamplitude as can be proved as follows.
Schematically the structure of the spinorial derivatives acting on external fields reads:
\beq
\int d^4\th\, \underbrace{D^2 \left( \dots \right)_{\text{left}}}_{\text{ odd $\#$ of $D^2$'s}}\, D^{\alpha}\, \F \bar D^2 \left( \Fb  \Fb \right) \, \bar D^{\dot \alpha}\, \underbrace{D^2 \left( \dots \right)_{\text{right}}}_{\text{ even $\#$ of $D^2$'s}}
\eeq
Focusing on the scalar component and recalling the equations of motion, we see that the $D^2 \bar D^2$ from the integration measure is not sufficient to cancel all the spinorial derivatives acting on superfields and potentially extracting their fermionic component.\\
For instance we may require the $D^2$ factor to act on $\bar D^2 \left( \Fb  \Fb \right)$ and the $\bar D^2$ on $D^2 \left( \dots \right)_{\text{left}}$ to put an even number of squared derivatives, but then $D^{\alpha}\, \F$ would survive, giving rise to a fermion as external state.
Therefore we conclude that the only contribution to the completely scalar scattering process comes from the scalar box integral above.
All other pentagon diagrams evaluate in the same fashion.

\end{document}